**Adapting the Interrelated Two-way Clustering method for Quantitative Structure-Activity Relationship (QSAR) Modeling of a Diverse Set of Chemical Compounds**

Subhabrata Majumdar[1], Subhash C. Basak*[2], Gregory D. Grunwald[2]

[1]School of Statistics, University of Minnesota- Twin Cities, 224 Church Street SE, Minneapolis, MN 55455 USA
[2]Natural Resources Research Institute, University of Minnesota Duluth, 5013 Miller Trunk Highway, Duluth, MN 55811, USA
*Author to whom all correspondence should be addressed.

Abstract: Interrelated Two-way Clustering (ITC) is an unsupervised clustering method developed to divide samples into two groups in gene expression data obtained through microarrays, selecting important genes simultaneously in the process. This has been found to be a better approach than conventional clustering methods like K-means or self-organizing map for the scenarios when number of samples much smaller than number of variables ($n<<p$). In this paper we used the ITC approach for classification of a diverse set of 508 chemicals regarding mutagenicity. A large number of topological indices (TIs), 3-dimensional, and quantum chemical descriptors, as well as atom pairs (APs) have been used as explanatory variables. In this paper, ITC has been used only for predictor selection, after which ridge regression is employed to build the final predictive model. The proper leave-one-out (LOO) method of cross-validation in this scenario is to take as holdout each of the 508 compounds *before* predictor thinning and compare the predicted values with the experimental data. ITC based results obtained here are comparable to those developed earlier.

1. Introduction

Prediction of mutagenicity is important both for drug discovery and environmental protection. If the mutagenicity of drug candidates is detected in the discovery phase of drug development, that can lead to better and earlier decisions about the allocation of resources in the costly drug discovery process which currently needs US $400 million to 2 billion per drug [1,2]. Regulatory agencies like the United States Environmental Protection Agency (USEPA) [3],routinely assess chemicals for their potential mutagenicity for hazard estimation. In the area of human health risk assessment of chemicals, carcinogenicity and mutagenicity are two toxicologically important end points. Because testing of a large number of chemicals in the laboratory is prohibitively expensive, alternative approaches to the bioassay were designed based on the molecular structure of chemicals as cost effective alternatives [4] for the identification of mutagens. Therefore, one recent trend in mutagenicity prediction is the use of theoretically computed descriptors in the development of models [5-7].

Different methods, like multiple regression, fuzzy logic [8], neural networks [8,9], multistep models [10] have been used by various authors for the prediction of mutagenicity of both congeneric as well as structurally diverse sets of chemicals. All of these use easily calculated molecular descriptors, including topological indices and atom pairs (APs). One of the smaller sets of chemical mutagens studied exhaustively is the group of 95 aromatic and heteroaromatic amines originally collected by Debnath *et al*[11]. A large number of studies have been reported in the literature on the QSARs of this set of chemicals using different classes of molecular descriptors [12-16]. In two earlier papers, Basak *et al* [17] and Hawkins *et al* [18] developed predictive models for mutagenicity of both a congeneric set of 95 aromatic amines and a diverse set of 508 chemicals consisting of mutagens and non-mutagens. The descriptors used included topostructural (TS), topochemical (TC), three dimensional (3-D), and quantum chemical (QC) descriptors. In our recent quantitative structure-activity relationship (QSAR) studies, we observed that the addition of calculated atom pairs (APs) to the collection of explanatory variables enhanced the quality of the models [19, 20]. So, it was of interest to investigate whether the addition of APs to the set of numerical molecular descriptors could help us in developing better models for chemical mutagenicity estimation.

We calculated a total of 2,525 descriptors for the 508 chemicals. For the selection of important independent variables from this large pool, we adapted a clustering approach first proposed by Tang *et al* [18]. Originally applied to gene expression data obtained from microarrays, this unsupervised method, named Interrelated Two-way Clustering (ITC), iteratively selects important genes and classifies samples simultaneously. Here we substitute samples for compounds, and genes for predictors. Since in our case we have the number of predictors nearly 5 times the number of samples, it fits the $n << p$ scenario in the gene data for which it is found to perform better than K-means or Self-Organizing Maps (SOM) clustering methods [21]. After predictor selection through ITC, ridge regression (RR) was used to build the final QSAR model for classifying compounds as mutagen/non-mutagen. An important point to be noted here is the approach to cross-validation. If we first select important predictors and then use cross-validation to select the tuning parameter in RR, it actually uses information from the holdout compound to build the model. So it is imperative that for each holdout sample compound, we do predictor thinning using other compounds and then use RR to predict the class of this compound [22]. Following this, the effectiveness of predictor selection by ITC is compared with the results obtained by modeling and subsequent classification results on the same dataset by Hawkins *et al* [18].

2. Materials and Methods

2.1 The database

The data were taken from the CRC Handbook of Identified Carcinogens and Non-carcinogens [23]. The response variable is Ames mutagenicity (which is an accurate indicator of carcinogenicity [24]), the sample available being 508 compounds classified as not mutagenic (scored 0) or mutagenic (scored 1). The set of 508 is comprised of 256 mutagens and 252 non-mutagens. Table 1 below gives an idea regarding the diversity of the chemicals in this database in terms of chemical types and functional groups.

<Insert Table 1 here>

2.2 Calculation of Descriptors

Software packages including POLLY v.2.3 [25], Sybyl v.6.2 [26], MOPAC v 6.00 [27] and Molconn-Z[28] were used to calculate descriptors, based solely on chemical structure. The triplet indices were calculated by in-house software developed by Basak *et al* [29] which can calculate descriptors formulated by Filip *et al* [30].Atom pairs are calculated by the software APProbe [31] which calculated APs following the method of Carhart *et al* [32]. The descriptors can be classified according to their complexity and demand on computational resources. The topostructural indices make up the simplest descriptor class, encoding information related solely to the connectedness of the atoms within a molecule. The topochemical descriptors are more complex, encoding not only information related to molecular topology but also information on atom and bond types. The geometric descriptors encode three-dimensional aspects of molecular structure; and the most complex and computationally demanding quantum chemical descriptors are based on the electronic aspects of molecular structure. Table 2 gives the symbols and definition of the majority of descriptors, including geometrical and quantum chemical indices, used in this study. Atom pairs are not given in a table because of the large size. The values of all calculated TIs, 3-D descriptors, QC chemical indices, and APs are given in the supplementary material accompanying the manuscript.

<Insert Table 2 here>

## 2.3 Statistical Analysis

### 2.3.1 An overview of Interrelated Two-way Clustering (ITC)

*The Method:* This method of unsupervised analysis aims to simultaneously select important predictors as well as cluster samples into two different classes (e.g. diseased and control) in a *single iterative procedure*. For this, at first predictors are divided into different groups using some known classification method or practical considerations and then samples are classified for each predictor group independently. The idea is that if the predictors are important in class detection, the sample classifications will be identical for each predictor group. To achieve this, in each iterative step some predictors are eliminated so that sample classifications based on different predictor groups become more and more similar.

The details of the procedure are as described below:

a. *Pre-processing:* In gene expression data, the predictors (i.e. genes) having little contribution in determining the class of a sample exhibit very little change in intensity values across different samples. The pre-processing stage is used to do a preliminary filtering of such predictors. For this the predictor vectors are first normalized [33]:

$$w'_{ij} = \frac{w_{ij} - \mu_i}{\mu_i}, \text{ with } \mu_i = \frac{\sum_{j=1}^{m} w_{ij}}{m}$$

where $w'_{ij}$ and $w_{ij}$ are the normalized and crude intensity values of the *i*-th predictor in the *j*-th compound, respectively, $i=1,2...n$, $n$ being the number of predictors and $m$ the number of compounds. After this the slope of each predictor vector $\boldsymbol{g_i} = (w'_{i1}, ..., w'_{im})'$ with respect to the constant vector $\boldsymbol{E}_{n \times 1} = (1, ..., 1)'$ is calculated:

$$\cos(\theta) = \frac{<\boldsymbol{g_i}, \boldsymbol{E}>}{\|\boldsymbol{g_i}\| \cdot \|\boldsymbol{E}\|} = \frac{\sum_{j=1}^{m} e_j w'_{ij}}{\sqrt{\sum_{j=1}^{m} {w'_{ij}}^2} \sqrt{\sum_{j=1}^{m} e_j^2}}$$

where $\theta$ is the angle between the predictor vector and the constant vector. A value close to 1 indicates the predictor intensity vector does not vary much across samples, while a value close to 0 implies near orthogonality of these two vectors, thus much deviation of the predictor expression from constant behavior across samples. In Tang *et al* [21], a threshold of 0.9 was used to filter the normalized vectors in this step.

b. *Details of the iterative procedure:* Each iteration involves the following steps:

STEP 1- Clustering in predictor dimension: Predictors are clustered into *k* groups ($G_1,...,G_k$) using a standard clustering algorithm like K-means [34, pp. 461] or SOM [34, pp. 480].

STEP 2- Clustering in sample dimension: For each predictor group $G_i$, samples are clustered in two groups $S_{i,a}$ and $S_{i,b}$, as per most popular experimental conditions [35].

STEP 3- Combining the two clusterings: For each predictor group two clusters are obtained, so each sample can be in any one of the two clusters for each of the $k$ groups. Thus there can be $2^k$ possible groups of samples. For example, for $k = 2$, the 4 groups will be:

$$C_1 = S_{1,a} \cap S_{2,a} \qquad C_2 = S_{1,b} \cap S_{2,a} \qquad C_3 = S_{1,a} \cap S_{2,b} \qquad C_4 = S_{1,b} \cap S_{2,b}$$

STEP 4- Obtaining heterogeneous groups: From these $2^k$ sample groups we select $2^{k-1}$ heterogeneous groups $(C_s, C_t)$. These groups are selected in such a way that for all $u \in C_s$ and $v \in C_t$, if $u \in S_{i,p}$, $v \in S_{i,q}$ then $a \neq b$ for all $i$. In the above case for k = 2, $(C_1, C_4)$ and $(C_2, C_3)$ are two heterogeneous groups.

STEP 5- Sorting and reducing: For each heterogeneous group $(C_s, C_t)$ two patterns are introduced: (0,0, … ,0,1,1, … ,1) and (1,1, … ,1,0,0, … ,0), containing $|C_s|$ (= #samples in $C_s$) zeros (ones) followed by $|C_t|$ ones (zeros), respectively. Vector cosines with these two patterns are calculated for predictor vectors in this heterogeneous group. The cosine values are sorted in decreasing order, and for each pattern the top one third of the predictors are kept. By merging the two sequences for the two complementary patterns we obtain a set of predictors which is reduced by at least one thirds from the predictors the iteration was started with.

Similar sets of predictors are generated for all the heterogeneous groups. Now, to select the final sequence of predictors for the iteration, leave-one-out cross validation is performed, i.e. for each heterogeneous group, select a sample, use the remaining samples to select important predictors, and use these predictors to predict the class of the withheld sample. This is repeated for all samples, and finally a cumulative error rate is obtained for each heterogeneous group. The group that has the lowest error rate has its corresponding reduced sequence of predictors selected as the set of predictors for next iteration.

c. *Termination condition:* We define the term *Occupancy ratio* as:

$$Occratio = Max \frac{|C_s + C_t|}{m}$$

where the maximum is taken over all heterogeneous groups $(C_s, C_t)$. When the predictor clustering results based on the predictor groups are the same, one of the heterogeneous groups will contain all the samples, thus the occ-ratio will be 1. Then the reduced predictor sequence obtained in STEP 5 will be good for sample clustering. Of course, this optimal condition is hard to reach in practical situations, so a threshold value of 0.9 is used as the termination condition for iterations. To include the cases where the occ-ratio value cannot reach the threshold after many iterations, yet the remaining number of predictors becomes too small, a specific threshold of 100 predictors is used as an alternate termination condition.

*Adapting the ITC in QSAR scenario:* In our case, predictors take the place of genes, and samples are substituted by sample compounds. Since number of predictors is much higher than the number of compounds, this is also a case of the *n < p* scenario in which the ITC was originally applied. The ITC is applied here in the following way:

   a. We already have the classification of predictors: TS/TC/3D/QC/AP; so the clustering in gene direction was not needed.
   b. For each predictor group, sample compounds were clustered using K-means ($k = 2$).

2.3.2 Transformation of predictors

Before applying any statistical procedure, the data containing selected predictors are transformed. Because differences in magnitudes across predictors might not be of the same order, each entry *x* in the data is transformed as *x' = log(x+C)* where *C = 1* when *x > 0*, and otherwise is the negative of the largest integer less than *x*. For example, if *x = -1.7617* then we take *C = 2*. After this transformation on the explanatory variables, we center and scale the response and explanatory variables by subtracting and dividing each entry by the mean and standard deviation of that column, respectively.

2.3.3 Ridge Regression

Ridge Regression (RR) [36, pp. 239] is used in place of OLS regression where significant correlation exists between different explanatory variables. For a given data (centered and scaled, no intercept term), with **X** being the matrix of predictors and **Y** the vector of responses among samples, the vector **b** of estimates for coefficients obtained by RR is given by

$$\boldsymbol{b} = (\boldsymbol{X}'\boldsymbol{X} + k\boldsymbol{I})^{-1}\boldsymbol{X}'\boldsymbol{Y}$$

Where *k > 0* is the ridge constant. A value of *k = 0* corresponds to OLS regression, while as *k* grows to infinity, the RR estimates shrink towards 0.

Methods of choosing *k* suitably include using leave-one-out prediction sum of squares (PRESS) statistic and Generalized Cross-Validation (GCV) [36, pp. 253]. In each of these the fitted cross-validation score is calculated for a range of values, and we choose as *k* the value for which this score is minimum. Finally the predictive ability of the model is assessed by cross-validated classification score, which is obtained the following way:

   a. Remove one compound from the data set. Fit a RR model with the rest of the compounds by choosing an optimal *k* and obtaining the corresponding vector of coefficients.
   b. Obtain the predicted response for the holdout compound by multiplying this vector of coefficients with the predictor vector of this compound. If the value obtained is greater than a previously fixed cutoff value, say, *c*, then take the predicted mutagenicity score as 1, otherwise take it 0.

c.  Continue for all the sample compounds. Now by comparing the true and predicted mutagenicity classifications we obtain the cross-validated classification score.

2.3.4 Naïve and proper cross validation

When the setup includes a step to select important variables before performing the model-building, it is imperative that the cross-validation is performed at the proper stage. It must be ensured that no information from the holdout compound is used in any way for predicting its response through cross-validation. That is why first doing predictor selection and then using cross-validation to obtain the model is not the proper way to do cross-validation [22], because then the first step of thinning involves the holdout compound as well. So, in our situation we do the cross-validation by omitting each compound, separately perform the ITC clustering routine and choose the ridge constant $k$ for all of them, and then predict mutagenicity scores of the compounds.

3. Results and Discussion

The objectives of our work were two-fold:

a) To investigate how far the addition of AP descriptors to the set of TS+TC+3D+QC indices increases the quality of models for predicting mutagenicity of chemicals, and

b) To adapt the ITC method, originally developed for application in gene expression data analysis, in the selection of a subset of explanatory variables from a large pool of descriptors.

All the analyses were performed in MATLAB, version R2008a [37]. We followed a hierarchical approach by first performing RR on only TS and TC descriptors. RR results involving TS, TC, 3D and QC descriptors on the same set of chemicals were obtained by Hawkins *et al* [18] previously. As we can see in the results in Table 3, the predictive ability before and after including the 3D and QC descriptors is almost the same, with the TS+TC model having a loss of sensitivity (i.e. correct prediction percentage for mutagens) of 0.4 and the same specificity (i.e. correct prediction percentage for non-mutagens), translating to misclassification of only one mutagen compared to the TS+TC+3D+QC model. This is in line with earlier hierarchical QSAR (HiQSAR) studies of Basak *et al* [12,14,15,17,38,39] for various sets of physicochemical property, bioactivity, and toxicity data that 3-D and QC descriptors make very little or no improvement in model quality after the use of TS and TC descriptors.

Taking this into account, and also the fact that including 5 types of descriptors in the ITC algorithm would result in $2^4 = 16$ heterogeneous groups and thus significant increase of computational load, we decided to perform the ITC thinning on TS+TC+AP descriptors instead of the TS+TC+3D+QC+AP descriptors. In the first iteration, the occ-ratio reached 0.89 and the descriptors gave better predictive scores than by RR on the full TS+TC+3D+QC set. Going into the second iteration, the occ-ratio improved slightly to 0.9075, terminating the algorithm, but the number of predictors was diminished, resulting in significant decrease of predictive ability. Because of this the model built from the set of descriptors obtained after the first iteration was taken as final.

Cutoff for predicted mutagenicity was taken as $c = 0.5$ for all methods. The results are summarized in the table below

< Insert Table 3 here>

The set of predictors obtained after ITC thinning contained 57 topostructural, 101 topochemical and 45 atom-pair descriptors. Although this model contained fewer predictors, it gave a 2% increase of specificity than RR using first 4 types of descriptors. It is also to be noted that ITC analysis reported here did variable selection from a large pool of descriptors contained in the TS+TC+AP set of explanatory variables.

12 descriptors (6 topochemical, 6 atom-pair) were found to have /$t$/-ratios that are significant at 95% confidence level. Interestingly, no topostructural descriptors were among these 12. One possible interpretation could be that of the compound set containing a large variety of compounds of different structural classes and the structural information encoded by just the connectivity of atoms without any consideration of the atomic characters or bonding patterns were not enough to predict mutagenicity efficiently. Therefore all the 6 influential TIs were electrotopological state indices. The /$t$/-ratios and names and types of these descriptors are as in Table 4:

<Insert Table 4 here>

4. Conclusion

The main objective of this paper was to study the effectiveness of ITC method vis-à-vis RR technique in the prediction of mutagenicity/ non-mutagenicity of a diverse set of chemicals. Results show that ITC can do effective variable selection from a large pool of calculated descriptors. Predictive models developed from the ITC-derived descriptors compare reasonably well with those developed using the RR method. Further studies with other sets of bioactive chemicals, as well as employing cross-validation with a training set of 10-20% of samples to protect against overfitting and comparing with the results obtained by leave-one out CV are needed to characterize the relative effectiveness of these methods in QSAR development.

5. Acknowledgement

This is contribution number XXX from the Center for Water and the Environment of the Natural Resources Research Institute, University of Minnesota Duluth, USA. The authors would like to thank Professor Douglas Hawkins, School of Statistics, University of Minnesota, for his help in data analysis carried out in this paper. One of the authors (SBM) would like to thank Satadip Saha for helping in running the codes.

Table 1. Major chemical classes (not mutually exclusive) within the mutagen/non-mutagen database.

| Chemical class | Number of compounds |
|---|---|
| Aliphatic alkanes, alkenes, alkynes | 124 |
| Monocyclic compounds | 260 |
|     Monocyclic carbocycles | 186 |
|     Monocyclic heterocycles | 74 |
| Polycyclic compounds | 192 |
|     Polycyclic carbocycles | 119 |
|     Polycyclic heterocycles | 73 |
| Nitro compounds | 47 |
| Nitroso compounds | 30 |
| Alkyl halides | 55 |
| Alcohols, thiols | 93 |
| Ethers, sulfides | 38 |
| Ketones, ketenes, imines, quinones | 39 |
| Carboxylic acids, peroxy acids | 34 |
| Esters, lactones | 34 |
| Amides, imides, lactams | 36 |
| Carbamates, ureas, thioureas, guanidines | 41 |
| Amines, hydroxylamines | 143 |
| Hydrazines, hydrazides, hydrazones, traizines | 55 |
| Oxygenated sulfur and phosphorus | 53 |
| Epoxides, peroxides, aziridines | 25 |

Table 2. Symbols, definitions and classification of topological indices

| | Topostructural (TS) |
|---|---|
| $I_D^W$ | Information index for the magnitudes of distances between all possible pairs of vertices of a graph |
| $\overline{I}_D^W$ | Mean information index for the magnitude of distance |
| $W$ | Wiener index = half-sum of the off-diagonal elements of the distance matrix of a graph |
| $I^D$ | Degree complexity |
| $H^V$ | Graph vertex complexity |
| $H^D$ | Graph distance complexity |
| $\overline{IC}$ | Information content of the distance matrix partitioned by frequency of occurrences of distance $h$ |
| $M_1$ | A Zagreb group parameter = sum of square of degree over all vertices |
| $M_2$ | A Zagreb group parameter = sum of cross-product of degrees over all neighboring (connected) vertices |
| $^h\chi$ | Path connectivity index of order $h$ = 0-10 |
| $^h\chi_C$ | Cluster connectivity index of order $h$ = 3-6 |
| $^h\chi_{PC}$ | Path-cluster connectivity index of order $h$ = 4-6 |
| $^h\chi_{Ch}$ | Chain connectivity index of order $h$ = 3-10 |
| $P_h$ | Number of paths of length $h$ = 0-10 |
| $J$ | Balaban's $J$ index based on topological distance |
| $nrings$ | Number of rings in a graph |
| $ncirc$ | Number of circuits in a graph |
| $DN^2S_y$ | Triplet index from distance matrix, square of graph order, and distance sum; operation $y$ = 1-5 |
| $DN^21_y$ | Triplet index from distance matrix, square of graph order, and number 1; operation $y$ = 1-5 |
| $AS1_y$ | Triplet index from adjacency matrix, distance sum, and number 1; operation $y$ = 1-5 |
| $DS1_y$ | Triplet index from distance matrix, distance sum, and number 1; operation $y$ = 1-5 |
| $ASN_y$ | Triplet index from adjacency matrix, distance sum, and graph order; operation $y$ = 1-5 |
| $DSN_y$ | Triplet index from distance matrix, distance sum, and graph order; operation $y$ = 1-5 |
| $DN^2N_y$ | Triplet index from distance matrix, square of graph order, and graph order; operation $y$ = 1-5 |
| $ANS_y$ | Triplet index from adjacency matrix, graph order, and distance sum; operation $y$ = 1-5 |
| $AN1_y$ | Triplet index from adjacency matrix, graph order, and number 1; operation $y$ = 1-5 |
| $ANN_y$ | Triplet index from adjacency matrix, graph order, and graph order again; operation $y$ = 1-5 |
| $ASV_y$ | Triplet index from adjacency matrix, distance sum, and vertex degree; operation $y$ = 1-5 |
| $DSV_y$ | Triplet index from distance matrix, distance sum, and vertex degree; operation $y$ = 1-5 |
| $ANV_y$ | Triplet index from adjacency matrix, graph order, and vertex degree; operation $y = 1\text{-}5$ |
| $kp_0$ | Kappa zero |
| $kp_1\text{-}kp_3$ | Kappa simple indices |

| Topochemical (TC) | |
|---|---|
| O | Order of neighborhood when $IC_r$ reaches its maximum value for the hydrogen-filled graph |
| $O_{orb}$ | Order of neighborhood when $IC_r$ reaches its maximum value for the hydrogen-suppressed graph |
| $I_{ORB}$ | Information content or complexity of the hydrogen-suppressed graph at its maximum neighborhood of vertices |
| $IC_r$ | Mean information content or complexity of a graph based on the $r^{th}$ ($r = 0\text{-}6$) order neighborhood of vertices in a hydrogen-filled graph |
| $SIC_r$ | Structural information content for $r^{th}$ ($r = 0\text{-}6$) order neighborhood of vertices in a hydrogen-filled graph |
| $CIC_r$ | Complementary information content for $r^{th}$ ($r = 0\text{-}6$) order neighborhood of vertices in a hydrogen-filled graph |
| $^h\chi^b$ | Bond path connectivity index of order $h = 0\text{-}6$ |
| $^h\chi^b_C$ | Bond cluster connectivity index of order $h = 3\text{-}6$ |
| $^h\chi^b_{Ch}$ | Bond chain connectivity index of order $h = 3\text{-}6$ |
| $^h\chi^b_{PC}$ | Bond path-cluster connectivity index of order $h = 4\text{-}6$ |
| $^h\chi^v$ | Valence path connectivity index of order $h = 0\text{-}10$ |
| $^h\chi^v_C$ | Valence cluster connectivity index of order $h = 3\text{-}6$ |
| $^h\chi^v_{Ch}$ | Valence chain connectivity index of order $h = 3\text{-}10$ |
| $^h\chi^v_{PC}$ | Valence path-cluster connectivity index of order $h = 4\text{-}6$ |
| $J^B$ | Balaban's $J$ index based on bond types |
| $J^X$ | Balaban's $J$ index based on relative electronegativities |
| $J^Y$ | Balaban's $J$ index based on relative covalent radii |
| $AZV_y$ | Triplet index from adjacency matrix, atomic number, and vertex degree; operation $y = 1\text{-}5$ |
| $AZS_y$ | Triplet index from adjacency matrix, atomic number, and distance sum; operation $y = 1\text{-}5$ |
| $ASZ_y$ | Triplet index from adjacency matrix, distance sum, and atomic number; operation $y = 1\text{-}5$ |
| $AZN_y$ | Triplet index from adjacency matrix, atomic number, and graph order; operation $y = 1\text{-}5$ |
| $ANZ_y$ | Triplet index from adjacency matrix, graph order, and atomic number; operation $y = 1\text{-}5$ |
| $DSZ_y$ | Triplet index from distance matrix, distance sum, and atomic number; operation $y = 1\text{-}5$ |
| $DN^2Z_y$ | Triplet index from distance matrix, square of graph order, and atomic number; operation $y = 1\text{-}5$ |
| $nvx$ | Number of non-hydrogen atoms in a molecule |
| $nelem$ | Number of elements in a molecule |
| *fw* | Molecular weight |
| *si* | Shannon information index |
| *totop* | Total Topological Index *t* |
| *sumI* | Sum of the intrinsic state values *I* |
| *sumdelI* | Sum of delta-*I* values |
| *tets2* | Total topological state index based on electrotopological state indices |
| *phia* | Flexibility index ($kp_1$* $kp_2$/$nvx$) |
| *Idcbar* | Bonchev-Trinajstić information index |
| *IdC* | Bonchev-Trinajstić information index |
| *Wp* | Wienerp |
| *Pf* | Plattf |
| *Wt* | Total Wiener number |
| *knotp* | Difference of chi-cluster-3 and path/cluster-4 |
| *knotpv* | Valence difference of chi-cluster-3 and path/cluster-4 |
| *nclass* | Number of classes of topologically (symmetry) equivalent graph vertices |
| *NumHBd* | Number of hydrogen bond donors |
| *NumHBa* | Number of hydrogen bond acceptors |
| *SHCsats* | E-State of C *sp$^3$* bonded to other saturated C atoms |
| *SHCsatu* | E-State of C *sp$^3$* bonded to unsaturated C atoms |
| *SHvin* | E-State of C atoms in the vinyl group, *=CH-* |
| *SHtvin* | E-State of C atoms in the terminal vinyl group, *=CH$_2$* |
| *SHavin* | E-State of C atoms in the vinyl group, *=CH-*, bonded to an aromatic C |
| *SHarom* | E-State of C *sp$^2$* which are part of an aromatic system |
| *SHHBd* | Hydrogen bond donor index, sum of Hydrogen E-State values for *–OH*, *=NH*, *-NH$_2$*, *-NH-*,*-SH*, and *#CH* |
| *SHwHBd* | Weak hydrogen bond donor index, sum of *C-H* Hydrogen E-State values for hydrogen atoms on a C to which a F and/or Cl are also bonded |
| *SHHBa* | Hydrogen bond acceptor index, sum of the *E*-State values for *–OH*, *=NH*, *-NH$_2$*, *-NH-*, *>N*, *-O-*, *-S-*, along with *–F* and *–Cl* |
| *Qv* | General Polarity descriptor |
| *NHBint$_y$* | Count of potential internal hydrogen bonders ($y$ = 2-10) |
| *SHBinty* | E-State descriptors of potential internal hydrogen bond strength ($y$ =2-10) |
| *ka$_1$-ka$_3$* | Kappa alpha indices |
| | Electrotopological State index values for atom types:<br>   *SHsOH, SHdNH, SHsSH, SHsNH2, SHssNH, SHtCH, SHother, SHCHnX, HmaxGmax, Hmin, Gmin, Hmaxpos, Hminneg, SsLi, SssBe, Sssss, Bem, SssBH ,SsssB, SssssBm, SsCH3, SdCH2, SssCH2, StCH, SdsCH, SaaCH, SsssCH, SddC, StsC, SdssC, SaasC, SaaaC, SssssC, SsNH3p, SsNH2, SssNH2p, SdNH, SssNH, SaaNH, StN, SsssNHp, SdsN, SaaN, SsssN, SddsN, SaasN, SssssNp, SsOH, SdO, SssO, SaaO, SsF, SsSiH3, SssSiH2, SsssSiH, SssssSi, SsPH2, SssPH, SsssP, SdsssP, SssssssP, SsSH, SdS, SssS, SaaS, SdssS, SddssS, SssssssS, SsCl, SsGeH3, SssGeH2, SsssGeH, SssssGe, SsAsH2, SssAsH, SsssAs, SdsssAs, SssssssAs, SsSeH, SdSe, SssSe, SaaSe, SdssSe, SddssSe, SsBr, SsSnH3, SssSnH2, SsssSnH, SssssSn, SsI, SsPbH3, SssPbH2, SsssPbH, SssssPb*|

|  | Geometrical (3-D) |
| --- | --- |
| $^{3D}W$ | 3D Wiener number based on the hydrogen-suppressed geometric distance matrix |
| $^{3D}W_H$ | 3D Wiener number based on the hydrogen-filled geometric distance matrix |
| $V_W$ | Van der Waal's volume |
|  | Quantum Chemical (QC) |
| $E_{HOMO}$ | Energy of the highest occupied molecular orbital |
| $E_{HOMO-1}$ | Energy of the second highest occupied molecular |
| $E_{LUMO}$ | Energy of the lowest unoccupied molecular orbital |
| $E_{LUMO+1}$ | Energy of the second lowest unoccupied molecular orbital |
| $\Delta Hf$ | Heat of formation |
| $\mu$ | Dipole moment |

Table 3. Results of RR analyses of 508 mutagens and non-mutagens using calculated descriptors

| Type of predictors in model | Model description | No. of predictors | Type of cross validation | Correct classification % | Sensitivity | Specificity |
|---|---|---|---|---|---|---|
| TS+TC | Ridge regression without descriptor thinning | 298 | Leave-one-out CV | 76.97 | 83.98 | 69.84 |
| TS+TC+3D+QC | Ridge regression without descriptor thinning | 307 | Leave-one-out CV, done by Hawkins *et al* [18] | 77.17 | 84.38 | 69.84 |
| TS+TC+AP | RR with ITC thinning (after first iteration) | 203 | Two-deep CV | 78.35 | 84.38 | 72.22 |

Table 4. Descriptors with significant |*t*|-values from the RR model obtained using descriptors selected by ITC

| Descriptor name | \|*t*\|-ratio | Descriptor class |
|---|---|---|
| *SsCH3* | 3.7555 | TC |
| *O.X$_1$-3-O.X$_1$* | 3.4157 | AP |
| *NX$_3$-3-S.X$_1$* | 2.8252 | AP |
| *C.X$_3$-2-NX$_3$* | 2.8246 | AP |
| *SaaNH* | 2.7164 | TC |
| *SdS* | 2.4749 | TC |
| *CX$_1$-3-OX$_2$* | 2.4209 | AP |
| *OX$_2$-2-S..X$_4$* | 2.2023 | AP |
| *StsC* | 2.1987 | TC |
| *C.X$_3$-2-C.X$_3$* | 2.1623 | AP |
| *SsNH2* | 2.1488 | TC |
| *SDsssP* | 2.0329 | TC |